\documentclass[12pt]{iopart}
\usepackage{iopams}
\begin{document}
\title{A simple estimate of gravitational wave memory in binary black hole systems}

\author{David Garfinkle}

\address{Dept. of Physics, Oakland University, Rochester, MI 48309, USA}
\address{Michigan Center for Theoretical Physics, Randall Laboratory of Physics, University of Michigan, Ann Arbor, MI 48109-1120, USA}

\ead{garfinkl@oakland.edu}

\begin{abstract}

A simple estimate is given of gravitational wave memory for the inspiral and merger of a binary black hole system.  Here the memory is proportional to the total energy radiated and has a simple angular dependence.  Estimates of this sort might be helpful as a consistency check for numerical relativity memory waveforms.

\end{abstract}

\maketitle

\section{Introduction}

With the detection of gravitational waves from merging black holes\cite{gw2015} we have now entered the era of gravitational wave astronomy.  One aspect of gravitational waves that has not yet been detected is gravitational wave memory: a displacement of the components of the detector that persists even after the wave has passed.  Gravitational wave memory was first treated within the weak field approximation to general relativity by Zeldovich and Polnarev\cite{zeldovich}, and in the full nonlinear theory by Christodoulou\cite{christodoulou}.  For a recent review of memory see \cite{bgy}. As an overall change between the initial and final state of the detector, memory would seem to be an inherently low frequency signal, likely to be overwhelmed by the low frequency noise that is an inescapable aspect of ground based gravitational wave detectors.  However, for the purposes of detection it is perhaps better to think of memory as an aspect of the waveform, where the net displacement is attained with a specific rise time.  This memory waveform has been estimated using a semi-analytic approximation,\cite{favata1,favata2} calibrated using the results of the numerical simulations.  The waveforms of \cite{favata1,favata2} have been used as part of a suggested strategy\cite{chen} for eventually detecting memory by combining the results of many gravitational wave detections.  Unfortunately, the complications of the methods for estimating the memory waveform given in \cite{favata1,favata2} and for detecting it \cite{chen} tend to leave the impression that gravitational wave memory is a very complicated phenomenon.  However, though the rise time and rise profile of the memory may be complicated, the size of the memory is rather simple, and is expressed in terms of the radiated energy per unit solid angle in straightforward expressions in\cite{christodoulou} and \cite{bg}.  As we will show in this Note, in the case of binary black holes of comparable mass there is a further simplification that gives rise to a very simple expression for the overall memory.  Section 2 contains a derivation of this simple expression.  Section 3 generalizes this result to a slightly less simple expression for memory.  The conclusion considers the implications for detection of memory and for extraction of memory from the numerical simulations.

\section{A Simple Memory Estimate}

Consider two masses in free fall, at a distance $r$ from a source of gravitational waves, and with an intial separation of $d^a$ before the gravitational wave passes.  Then after the wave has passed the separation will have changed by $\Delta {d^a}$ given by
\begin{equation}
{\Delta {d^a}} = {\frac {-1} r} {{m^a}_b} {d^b}
\end{equation}
where ${m^a}_b$ is the memory tensor.  We will find it useful to expand the memory tensor in spherical harmonics:
\begin{equation}
{m_{ab}} = {\sum _{\ell > 1}} {b_{\ell m}} \left ( {D_a}{D_b} {Y_{\ell m}} - {\textstyle {\frac 1 2}} {h_{ab}} {D_c}{D^c} {Y_{\ell m}} \right ) \; \; \; .
\label{memorysum}
\end{equation}
Here the tensor indices are on the unit two-sphere, with $h_{ab}$ and $D_a$ being the two-sphere metric and derivative operator, and $Y_{\ell m}$ being spherical harmonics.  
Let $F$ be the energy per unit solid angle radiated to null infinity.  Then the memory tensor is determined by $F$. In \cite{bg} an expression was given for the spherical harmonic coefficients $b_{\ell m}$ in terms of $F$ as follows:
\begin{equation}
{b_{\ell m}} = {\frac {-16\pi} {(\ell -1) \ell (\ell+1)(\ell+2)}} \; {\int d \Omega} \, {Y^* _{\ell m}} \, F
\label{blm}
\end{equation}
Here the integral is over the 2-sphere, with $d \Omega$ the usual 2-sphere volume element.  As argued in \cite{bg}, due to the large power 
of $\ell$ in the denominator of eqn. (\ref{blm}), we expect the sum in eqn. (\ref{memorysum}) to be dominated by the lowest $\ell$ terms, that is the terms with $\ell=2$.  

Now we consider the form of $F$ for the case of binary black hole mergers of equal mass.  As shown in \cite{fa}, for this case the Weyl component $\Psi_4$ is predominantly given by a combination of the spin weighted spherical harmonics 
$_{-2} Y_{22}$ and $_{-2} Y_{2-2}$, whose explicit expression is
\begin{eqnarray}
{_{-2} Y_{22}} = {\frac 1 8} {\sqrt {\frac 5 \pi}} \, {{(1+\cos \theta )}^2} \, {e^{2 i \phi}}
\nonumber
\\
{_{-2} Y_{2-2}} = {\frac 1 8} {\sqrt {\frac 5 \pi}} \, {{(1-\cos \theta )}^2} \, {e^{-2 i \phi}}
\label{spw2}
\end{eqnarray}
(Here the motion of the binary system is confined to a plane, and $\theta$ is the polar angle with respect to the plane, {\it i.e.} the plane is at $\theta =\pi/2$).
It then follows that the News function $N$ is also a combination of these same spin weighted spherical harmonics.  However, the power radiated per unit solid angle is given by $dP/d\Omega = {|N|}^2$, and the energy radiated per unit solid angle is given by $F = {\int _{-\infty} ^\infty} \, dt \, dP/d\Omega$.  It then follows using eqn. (\ref{spw2}) that $F$ takes the form
\begin{eqnarray}
F = \alpha ({\cos ^4} \theta + 6 {\cos ^2} \theta + 1 ) + \beta \cos \theta (1 + {\cos ^2} \theta ) 
\nonumber
\\
+ {\sin^4} \theta ({\gamma _1} \cos 4 \phi + {\gamma _2} \sin 4 \phi )
\label{F1}
\end{eqnarray}  
where $\alpha, \, \beta, {\gamma _1},$ and $\gamma _2$ are constants.  However, we now argue that $\gamma _1$ and $\gamma _2$ are negligible: in $dP/d\Omega$ the corresponding terms are not negligible, however as the binary black hole system rotates, so does this part of the power radiated, and therefore we expect that this term in $dP/d\Omega$
averages out to zero when integrated over time and therefore gives negligible contribution to $F$.  The remaining coefficients in eqn. (\ref{F1}) are determined by the total radiated energy $E$ and total radiated momentum ${p_z}{\hat z}$ as follows: we have $ E = \int F \, d \Omega$ and therefore using eqn. (\ref{F1}) we find 
$E = 64 \pi \alpha /5$ and thus $\alpha = 5 E/(64 \pi)$.  Similarly, we have ${p_z} = \int F \, \cos \theta \, d \Omega$ and therefore using eqn. (\ref{F1}) we find ${p_z} = 32\pi \beta/15$ and thus $\beta = 15 {p_z}/(32 \pi)$.  We 
therefore have
\begin{equation}
F = {\frac 5 {64\pi}} \, \left [ E ({\cos ^4} \theta + 6 {\cos ^2} \theta + 1 ) \; 
+ \; 6 {p_z} \cos \theta (1 + {\cos ^2} \theta ) \right ]
\label{F2}
\end{equation}
It will be helpful to express $F$ in terms of spherical harmonics.  Note that this expansion requires only spherical harmonics with $m=0$ and $\ell \le 4$.  Defining $x \equiv \cos \theta$ these spherical harmonics are
\begin{eqnarray}
{Y_{00}} = {\frac 1 {\sqrt {4\pi}}} 
\nonumber
\\
{Y_{10}} = {\sqrt {\frac 3 {4\pi} }} \, x
\nonumber
\\
{Y_{20}} = {\sqrt {\frac 5 {4\pi} }} \, {\textstyle {\frac 1 2}} (3 {x^2} - 1)
\nonumber
\\
{Y_{30}} = {\sqrt {\frac 7 {4\pi} }} \, {\textstyle {\frac 1 2}} (5 {x^3} - 3 x)
\nonumber
\\
{Y_{40}} = {\sqrt {\frac 9 {4\pi} }} \, {\textstyle {\frac 1 8}} (35 {x^4} - 30 {x^2} + 3)
\label{sph}
\end{eqnarray}
Using eqn. (\ref{sph}) in eqn. (\ref{F2}), some straightforward algebra yields
\begin{equation}
F = {\frac 1 {28 {\sqrt \pi}}} \, \left [ E \left ( {\frac 1 3} {Y_{40}} + 4 {\sqrt 5} {Y_{20}} + 14 {Y_{00}} \right )
+ {\frac {21} 2} \, {p_z} \left ( {\frac 1 {\sqrt 7}} {Y_{30}} + {\frac 4 {\sqrt 3}} {Y_{10}} \right ) \right ]
\label{F3}
\end{equation}

We now use eqn. (\ref{F3}) to find the memory.  In this section we use the approximation of only calculating the terms with $\ell=2$.  Note that this means that the memory will depend only on the total radiated energy $E$.  Also note that even if we had not argued that $\gamma_1$ and $\gamma_2$ are negligible, they would contribute to memory only through $\ell=4$ coefficients and therefore could be neglected in this approximation in which we keep only the $\ell=2$ terms.  Thus the only coefficient that we need to calculate is $b_{20}$.  Using eqn. (\ref{F3}) and the orthonormality of the spherical harmonics in
eqn. (\ref{blm}) we find
\begin{equation}
{b_{20}} = {\frac {-2 {\sqrt{5\pi}}} {21}} \, E
\label{b20}
\end{equation}
However, for any function $f(x)$ (where $x\equiv \cos \theta$) it follows from
standard properties of the 2-sphere metric and derivative operator that
\begin{equation}
{D_a}{D_b} f - {\textstyle {\frac 1 2}} {h_{ab}} {D_c}{D^c} f = {\textstyle {\frac 1 2}} \, {f ''}(x)
{\sin ^2} \theta ( {\theta _a}{\theta _b} - {\phi _a}{\phi _b})
\label{tensorf}
\end{equation}
Here $\theta _a$ and $\phi _a$ are respectively unit vectors in the $\theta$ and $\phi$ directions.
Then using eqn. (\ref{sph}) in eqn. (\ref{tensorf}) we find
\begin{equation}
{D_a}{D_b} {Y_{20}} - {\textstyle {\frac 1 2}} {h_{ab}} {D_c}{D^c} {Y_{20}} = {\frac 3 4} {\sqrt {\frac 5 \pi}}
{\sin ^2} \theta ( {\theta _a}{\theta _b} - {\phi _a}{\phi _b})
\label{tensorY20}
\end{equation}
Finally using eqns. (\ref{b20}) and (\ref{tensorY20}) in eqn. (\ref{memorysum}) we obtain
\begin{equation}
{m_{ab}} = - {\frac 5 {14}} E \, {\sin ^2} \theta \, ( {\theta _a}{\theta _b} - {\phi _a}{\phi _b})
\end{equation}
and thus the permanent displacement is 
\begin{equation}
\Delta {d^a} =  {\frac 5 {14}} \, {\frac E r} \, {\sin ^2} \theta \, ( {\theta ^a}{\theta _b} - {\phi ^a}{\phi _b})
\; {d^b}
\label{result}
\end{equation}

\section{A Slightly Less Simple Memory Estimate}

How good is the approximation used to derive eqn. (\ref{result})?  One way to address this issue is to note that in applying eqn. (\ref{memorysum}) we have neglected all terms with $\ell >2$.  Suppose that we still use the expression for $F$ in eqn. (\ref{F3}), but now no longer truncate the series at $\ell =2$.  How much will the result then differ from that of eqn. (\ref{result})?  For simplicity, we will only consider the case of no black hole kick, that is the case where ${p_z}=0$.  It then follows from eqn. (\ref{F3}) that the only term we have neglected is the one with $\ell=4$ and $m=0$.  From eqn. (\ref{F3}) and the orthornomality of the spherical harmonics we obtain
\begin{equation}
{b_{40}} = - {\frac {\sqrt \pi} {1890}} E
\label{b40}
\end{equation}  
Then using eqn. (\ref{sph}) in eqn. (\ref{tensorf}) we find
\begin{equation}
{D_a}{D_b} {Y_{40}} - {\textstyle {\frac 1 2}} {h_{ab}} {D_c}{D^c} {Y_{40}} = {\frac {45} {8{\sqrt \pi}}} (7 {\cos ^2}\theta - 1 ) 
{\sin ^2} \theta ( {\theta _a}{\theta _b} - {\phi _a}{\phi _b})
\label{tensorY40}
\end{equation}
Now using eqns. (\ref{b20}), (\ref{tensorY20}), (\ref{b40}), and (\ref{tensorY40}) in eqn. (\ref{memorysum}) we obtain a more accurate (though slightly less simple) formula for the memory tensor 
\begin{equation}
{m_{ab}} = - {\frac E {48}} (17 + {\cos^2} \theta) \, {\sin ^2} \theta \, ( {\theta _a}{\theta _b} - {\phi _a}{\phi _b})  
\end{equation}
and thus a more accurate (though slightly less simple) expression for the permanent displacement
\begin{equation}
\Delta {d^a} =   {\frac E {48 r}} \, (17 + {\cos ^2} \theta ){\sin ^2} \theta \, ( {\theta ^a}{\theta _b} - {\phi ^a}{\phi _b}) \; {d^b}
\label{result2}
\end{equation}
Eqn. (\ref{result2}) was derived by Favata in the context of the Post Newtonian approximation (see eqn. (5) of \cite{favata3}).

How does the result of eqn. (\ref{result2}) compare to the simple formula of eqn. (\ref{result})?  To make the comparison, note that eqn. (\ref{result2}) can be written as  
\begin{equation}
\Delta {d^a} =  {\frac 5 {14}} \, {\frac E r} \, {\sin ^2} \theta \, ( {\theta ^a}{\theta _b} - {\phi ^a}{\phi _b})
\; {d^b} \; \left [ {\frac {119} {120}} \left ( 1 + {\frac {{\cos ^2}\theta} {17}} \right ) \right ] 
\label{result3}
\end{equation}
Which is simply the expression of eqn. (\ref{result}) multiplied by the quantity in square brackets.  However, the quantity in square brackets is close to 1 for all angles, having a minimum value of approximately 0.992 (attained at $\theta =\pi/2$) and a maximum value of 1.05 (attained at $\theta =0$ or $\theta = \pi$).  Thus the result of eqn. (\ref{result}) is an excellent approximation to the result of eqn. (\ref{result2}).  

\section{Conclusions}

It is instructive to apply the simple estimate of memory given in eqn. (\ref{result}) to the case of GW150914.\cite{gw2015,gw2015p}  
In this case we have 
$E = 3.0 \, {M_{\odot}} = 4.4 \, {\rm km}, \; r = 410 \, {\rm Mpc} = 1.3 \times {{10}^{22}} \, {\rm km}, \; 
\theta = {{150}^\circ}$.  We therefore obtain
\begin{equation}
{\frac 5 {14}} \, {\frac E r} \, {\sin ^2} \theta = 3.0 \times {{10}^{-23}}
\end{equation}
Note that the strain signal for GW150914 has a maximum amplitude of $1.0 \times {{10}^{-21}}$.  Therefore we find that the memory signal is approximately $ 3\%$ of the maximum signal.  Note too that there is a factor of 1/4 due to the 
${\sin^2}\theta$.  For a system seen edge-on (i.e. with $\theta = {{90}^\circ}$) we would expect a memory signal of approximately $12\%$ of the maximum signal.  As argued in \cite{chen}, a signal of this size is unlikely to be detectable in a single event, but should be detectable by the ``stacking'' of multiple events.  

One would like the detection to occur through matching to a waveform generated by numerical simulations, so we next turn to the challenges involved in extracting the memory waveform from the simulations.  Memory is found by taking the dominant $r^{-1}$ piece of the Weyl tensor and integrating twice with respect to time.  The challenges to implementing this procedure consist of (i) the difficulty of picking out the 
dominant $r^{-1}$ piece when the extraction is done at finite radius, and (ii) the fact that in integrating twice with respect to time, even a small error in the integrand can lead to an error in the second integral that grows linearly with respect to time.  These challenges are addressed in \cite{pollney} where Cauchy-Characteristic matching is used to extract the $r^{-1}$ piece of the Weyl tensor, and matching to a post-Newtonian waveform at early times is done to make the double integral with respect to time accurate.  As a consistency check, one can compare the final amplitude of the memory waveform of the numerical simulations to the memory calculated by other means (such as eqn. (\ref{result}) of this Note).  In \cite{pollney} a consistency check of this sort is performed by comparing to the formula given in eqn. (5) of \cite{favata3}.  More generally, the method of \cite{pollney} also extracts the News tensor (which only takes one integration with respect to time); and the News tensor can be used to find the energy radiated per unit solid angle, which in turn can, using the formulas of \cite{christodoulou} or \cite{bg}, be used to calculate the Christodoulou memory.  Note however that in cases where there is a black hole kick, one must also include the ordinary memory of the recoiling black hole.\cite{bg,bgtw}  In this way, memory estimates (including the simple memory approximation of this Note) provide a useful consistency check for numerical memory waveforms, which in turn could be helpful in the eventual detection of gravitational wave memory.     
  
\section*{Acknowledgements}
I would like to thank Lydia Bieri, Frans Pretorius, Bob Wald, Bernard Whiting, and Nico Yunes for helpful discussions.  This work was supported by NSF grants PHY-1205202 and PHY-1505565 to Oakland University.

\end{document}